\documentclass[runningheads]{llncs}
\usepackage[T1]{fontenc}
\usepackage[x11names]{xcolor}
\usepackage{lscape}
\usepackage{graphicx}
\usepackage{booktabs}
\usepackage{colortbl}
\usepackage{pifont}
\usepackage{amssymb}
\usepackage{hyperref}
\usepackage{tablefootnote}
\usepackage{tikz}
\usepackage{url}
\usepackage{amsmath}
\usepackage{enumitem}
\usepackage{multirow}
\usepackage{longtable}
\usepackage{booktabs}
\usepackage{silence}
\usepackage[most]{tcolorbox}
\usepackage{tabularray}
\usepackage{colortbl}
\usepackage{hhline}
\usepackage{pdfpages}
\usepackage{algorithm}
\usepackage{siunitx}
\usepackage{mdframed}
\usepackage{listings}
\usepackage{nicematrix}

\usepackage{algpseudocode}  

\usepackage{ragged2e}

\usepackage{array}       
\usepackage{makecell}    
\usepackage{orcidlink}

\definecolor{headerColor}{RGB}{52, 73, 94}  
\definecolor{rowColor1}{RGB}{240, 248, 255} 
\definecolor{rowColor2}{RGB}{255,250,250} 

\definecolor{boxFrameColor}{RGB}{52, 73, 94}      
\definecolor{boxBackColor}{RGB}{255,250,250}    
\definecolor{titleBackColor}{RGB}{52, 73, 94}     
\definecolor{separatorColor}{RGB}{70, 130, 180}   

\definecolor{improvementGreen}{RGB}{0,128,0}
\definecolor{declineRed}{RGB}{220,20,60}
\definecolor{mongoBlue}{RGB}{230,240,255} 
\definecolor{tomcatColor}{RGB}{255,250,250} 

\newcommand{\xmark}{\ding{53}}  

\setlist[itemize]{topsep=1.65mm}

\usepackage{hyperref}

\usepackage{array} 
\newcolumntype{P}[1]{>{\centering\arraybackslash}p{#1}}
\newcolumntype{M}[1]{>{\centering\arraybackslash}m{#1}}

\usepackage{colortbl}

\tcbset{
    enhanced,
    breakable,
    listing only,
    fonttitle=\bfseries\color{white},
    colframe=boxFrameColor,
    colback=boxBackColor,
    coltitle=white,
    colbacktitle=titleBackColor,
    boxrule=2pt,                 
    titlerule=0.5pt,           
    arc=0mm,                     
    outer arc=0mm,                
    sharp corners,                 
    left=3mm,
    right=3mm,
    top=2mm,
    bottom=2mm,
    toptitle=0mm,
    bottomtitle=0mm,
    listing options={
        baselinestretch=1.2,
        fontsize=\footnotesize
    }
}

\newtcolorbox[auto counter]{promptbox}[2][]{
    listingboxes,
    title={Listing~\thetcbcounter: #2},
    halign title=center,
    fonttitle=\footnotesize\bfseries\color{white},
    width=4.8in,
    #1
}

\tcbset{
    listingboxes/.style={
    enhanced,
    breakable,
    listing only,
    fonttitle=\footnotesize\bfseries\color{white},
    colframe=boxFrameColor,
    colback=boxBackColor,
    coltitle=white,
    colbacktitle=titleBackColor,
    arc=0mm,                     
    outer arc=0mm,                
    sharp corners,                 
    left=4mm,
    right=3mm,
    top=0mm,
    bottom=2mm,
    titlerule=0pt,
    boxrule=2pt,
    toptitle=0mm,
    bottomtitle=0mm,
    boxsep=0pt, 
    titlerule=0pt,
    boxrule=2pt,
    listing options={
        baselinestretch=1.2,
        fontsize=\footnotesize
    },
    overlay={%
        \begin{tcbclipinterior}
            \fill[boxFrameColor!10] (frame.south west) rectangle ([xshift=3.5mm]frame.north west);
            \fill[boxFrameColor!10] (frame.south west) rectangle ([yshift=2mm]frame.south east);
            \fill[boxFrameColor!10] ([yshift=-4.6mm]frame.north west) rectangle (frame.north east);
            \fill[boxFrameColor!10] ([xshift=-2mm]frame.south east) rectangle (frame.north east);
        \end{tcbclipinterior}
    }
  }
}

\definecolor{lineNumColor}{RGB}{255,69,0}

\usepackage{listings}

\algrenewcommand\alglinenumber[1]{%
    \textcolor{lineNumColor}{\textbf{#1:}}%
}

\tcbset{highlight math style={colback=Ivory1!0, colframe=Snow1!0, boxrule=0.5pt, coltext=black}}

\definecolor{customblue}{HTML}{330099} 

\let\oldbibliography\thebibliography
\renewcommand{\thebibliography}[1]{%
  \oldbibliography{#1}%
  \setlength{\itemsep}{0pt}%
  \setlength{\parskip}{-1pt}%
}

\renewcommand{\footnotesize}{\scriptsize}

\vbadness=\maxdimen

\graphicspath{ {./images/} }


\title{Evaluating and Preventing Security Smells in AI-Generated Ansible Code}
\author{Pandu Ranga Reddy Konala\and
        Vimal Kumar \and
        David Bainbridge \and
        Junaid Haseeb}
\authorrunning{P. R. R. Konala~\emph{et al.}}

\titlerunning{Evaluating and Preventing Security Smells in AI-Generated Ansible Code}
\institute{School of Computing and Mathematical Sciences,\\
           University of Waikato,\\
           Hamilton, New Zealand 3240\\
           \email{\{pkonala, vkumar, davidb, jhaseeb\}@waikato.ac.nz}
           }

\begin{document}
\setlength{\hsize}{\hsize}
\justifying
\maketitle
\begin{abstract}
AI coding assistants generate Infrastructure as Code, yet no work has examined whether this code meets security requirements. This matters because security smells in infrastructure code propagate to deployed systems, producing infrastructure that is insecure and untrustworthy. We evaluate 16 AI models generating Ansible roles for Apache Tomcat v10 and MongoDB v7, analysing 278 ansible roles against CIS benchmarks. Without security guidance, all 16 AI models produced code containing security smells, resulting in vulnerable infrastructure that fails compliance verification and underperforms code written by human developers. We introduce an approach integrating Ansible best practices and CIS benchmarks into prompts through an extended CO-STAR framework, enabling security smell prevention during synthesis rather than detection after deployment. When this approach is applied, 4 out of 16 models generate compliant code, with the leading model achieving 95\%--100\% CIS compliance, a fourfold improvement over humans at 23\%--43\%, with overall code quality improving by 19\%--49\%. The remaining 12 models fail not because they cannot generate code but because they cannot follow instructions with multiple constraints. For capable models, the approach requires no retraining and can be adopted through system prompts.
\end{abstract}

\section{Introduction}
Infrastructure as Code (IaC) automates system configuration through specifications that replace manual provisioning with version-controlled deployments~\cite{KiefMorris}. Configuration management tools including Ansible, Terraform, and Puppet translate these specifications into deployed infrastructure. This automation introduces security risks, however. Security smells in infrastructure code, including hardcoded credentials, excessive permissions, and missing access controls, propagate directly to deployed systems, with the resulting infrastructure insecure and untrustworthy. Such infrastructure exposes organisations to data breaches, service disruptions, and regulatory penalties. In regulated industries, production deployments must satisfy compliance frameworks such as CIS Benchmarks\footnote{Center for Internet Security, ``CIS Critical Security Controls, Version 8,'' 2024, \href{https://www.cisecurity.org/controls/cis-controls-list/}{https://www.cisecurity.org/controls/cis-controls-list/}.} and DISA STIGs.\footnote{Defense Information Systems Agency, ``Security Technical Implementation Guide (STIG),'' 2025, \href{https://public.cyber.mil/stigs/downloads/}{https://public.cyber.mil/stigs/downloads/}.} Code containing security smells fails compliance verification and cannot receive deployment authorisation.

AI coding assistants compound these security risks. Tools such as Cursor,\footnote{Cursor AI Inc., ``Cursor,'' 2024, \href{https://cursor.sh}{https://cursor.sh}.} Bolt\footnote{StackBlitz Labs, ``Bolt: AI-Powered Full-Stack Web Development in the Browser,'' 2025, \href{https://github.com/stackblitz-labs/bolt.diy}{https://github.com/stackblitz-labs/bolt.diy}.} and others generate IaC from natural language descriptions using foundational AI models such as OpenAI's GPT-4o and Anthropic's Claude Opus 4 as their knowledge bases. Studies report that 45\%--62\% of AI-generated code contains vulnerabilities~\cite{csa2025aicode,veracode2025genai}. It is also reported that IaC misconfigurations account for 68\% of cloud security incidents~\cite{cloudpso2025misconfig}. Infrastructure misconfigurations can pose a greater security threat than application vulnerabilities because they become attack vectors immediately upon deployment.

Existing approaches to IaC security rely on detection rather than prevention, with static analysis tools identifying security smells after code enters repositories~\cite{pandusok}. This detection paradigm is inadequate for AI-generated code because each refinement cycle incurs time and computational costs while requiring security expertise for remediation. More fundamentally, detection cannot prevent security smells from entering codebases. Research on IaC compliance remains limited to frameworks without validation~\cite{Falazi2022_IaCComplianceManagement,Imperial2025_StandardizingIntelligence}, while studies on AI code security focus on application code rather than infrastructure~\cite{ji2024cybersecurity,tihanyi2024secure}. No known work has examined the security properties of AI-generated infrastructure code or whether security requirements can be embedded into the code generation process.

We investigate two questions: (\textbf{RQ1}) \textit{What is the default security posture of AI-generated infrastructure code and does it provide a viable foundation for achieving regulatory compliance?} (\textbf{RQ2}) \textit{Can security-compliance requirements be embedded into the generation process itself, eliminating the need for iterative post-generation remediation?} We evaluate 16 foundational AI models generating Ansible code for Apache Tomcat and MongoDB deployments, assessing 278 roles against CIS benchmarks through both pre-deployment code quality analysis and post-deployment compliance verification. Our contributions are:
\begin{itemize}
\item We evaluated 16 foundational AI models generating Ansible code, quantifying default security behaviour and standards awareness across open-source and closed-source implementations.
\item We extended the CO-STAR prompting framework to integrate Ansible best practices\footnote{Ansible Project, ``Ansible Best Practices -- Ansible Documentation,'' 2019, \href{https://docs.ansible.com/ansible/2.8/user_guide/playbooks_best_practices.html}{https://docs.ansible.com/ansible/2.8/user\_guide/playbooks\_best\_practices.html}.} and CIS benchmarks as generation constraints, enabling security smell prevention during synthesis for capable models without AI model retraining.
\end{itemize}
\section{Background}  
\label{cis_discussion}
Production deployments in regulated industries must satisfy compliance frameworks that establish security objectives for data protection and system hardening. These frameworks specify requirements without prescribing implementations, requiring organisations to translate security objectives into configurations including file permissions, authentication mechanisms, and encryption protocols. Technical implementation guides bridge this gap by providing technology-specific security guidelines.

The CIS publishes benchmarks for operating systems, databases, web servers, and cloud platforms. Each benchmark organises security controls into three Implementation Groups. IG1 defines basic cyber hygiene forming the minimum security baseline, while IG2 and IG3 add controls for environments facing sophisticated threats. Each control specifies security requirements, audit procedures, and remediation steps.

The DISA provides Security Technical Implementation Guides (STIGs) for government systems. While STIGs target federal deployments, CIS benchmarks achieve broader commercial adoption. U.S. federal frameworks such as FedRAMP \cite{fedramp} and DoD RMF~\cite{dod-rmf} mandate STIG compliance, while commercial frameworks such as PCI-DSS~\cite{pci-dss} and HIPAA~\cite{hipaa} accept CIS benchmarks as security implementation standards.

Compliance verification occurs through automated scanning tools. CIS-CAT Pro,\footnote{Center for Internet Security, ``CIS-CAT Pro: Official Configuration Assessment Tool for CIS Benchmarks,'' \href{https://www.cisecurity.org/cybersecurity-tools/cis-cat-pro}{https://www.cisecurity.org/cybersecurity-tools/cis-cat-pro}.} the official assessment tool for CIS benchmarks, performs security audits against benchmark controls, while SCAP Compliance Checker\footnote{Naval Information Warfare Center Atlantic, ``SCAP Compliance Checker,'' \href{https://www.niwcatlantic.navy.mil/Technology/SCAP/}{https://www.niwcatlantic.navy.mil/Technology/SCAP/}.} provides equivalent functionality for STIGs. These tools operate post-deployment, assessing running systems rather than the IaC that produced them. IaC generating non-compliant deployments cannot receive production authorisation, establishing security verification as a deployment gate. However, compliance with a specific framework does not guarantee security. Compliance controls are most effective when applied to code that already maintains a secure foundation, as regulatory requirements presuppose baseline security hygiene. Code quality thus serves as a prerequisite for effective compliance implementation.
\subsection{AI Code Generation}
While compliance frameworks define security and other requirements, AI models increasingly determine how IaC is generated. AI models generate code through next-token prediction, trained on corpora composed of of public repositories~\cite{vaswani2017attention}. These corpora contain code of varying security quality, from hardened production implementations to demonstration snippets with security smells~\cite{hajipour2023codelmsecbenchmarksystematicallyevaluating}. Training objectives reward syntactic validity and functional correctness without evaluating security properties, and models have limited mechanisms to distinguish secure configurations from IaC with security smells~\cite{cheng2025benchmarkingbrokendont,eriksson2025trust}.

AI coding assistants such as Cursor~\cite{cursor} and Bolt~\cite{BoltDIY} provide interfaces to these models, processing natural language prompts and returning generated code. These tools employ system prompts to define role, output format, and behavioural constraints~\cite{meta_llama4_system_prompt_2025,bolt_systemprompt}. Analysis of system prompts from open-source tools such as Bolt and Llama reveals no instructions regarding security practices. Models therefore tend to reproduce patterns from training data without reliably distinguishing secure configurations from vulnerable ones. This shifts the quality-compliance burden from tool to operator, requiring users to specify security requirements explicitly.
\subsection{Related Work}
Related work addresses compliance management frameworks, AI code security, and IaC security analysis, though none examines AI-generated infrastructure code for compliance. Compliance management frameworks for IaC remain theoretical. Falazi~\emph{et al.}~\cite{Falazi2022_IaCComplianceManagement} proposed a technology-agnostic compliance management framework addressing configuration drift, and Imperial~\emph{et al.}~\cite{Imperial2025_StandardizingIntelligence} examined generative AI alignment with regulatory frameworks, yet neither provides empirical validation. AI code security research demonstrates that models introduce vulnerabilities in practice. Tihanyi~\emph{et al.}~\cite{tihanyi2024secure} found 62\% of 331,000 AI-generated C programs contained vulnerabilities and Ji~\emph{et al.}~\cite{ji2024cybersecurity} reported 48\% of AI-generated code contains MITRE CWE Top 25 weaknesses. However, these studies focus on application code security rather than infrastructure security; whether similar patterns manifest in IaC was not examined.

Independent of AI-based code generation, IaC security research reveals deficiencies in human-written implementations. Rahman~\emph{et al.}~\cite{rahman2019security} identified seven security smell categories across 1,093 repositories. Similar studies employ detection-based approaches, identifying security violations after code enters repositories rather than preventing them during generation. Static security smell detection tools surveyed by Konala~\emph{et al.}~\cite{pandusok} assume human developers who implement security corrections iteratively. For AI-generated code, this approach provides no mechanism to prevent security smells during synthesis. No prior empirical work was found that examined the security properties of AI-generated infrastructure code or its viability for achieving compliance.
\section{Baseline Security Performance of AI Models}
To address \textbf{RQ1}, we evaluated foundational AI models generating Ansible roles for Apache Tomcat deployment without explicit security guidance. These outputs were compared against human-written implementations from community repositories, establishing baseline security performance and identifying deficiencies in AI-generated code.
\subsection{Evaluation Protocol}
To measure default security behaviour, we evaluated 16 foundational AI models using identical zero-shot prompts. We assessed the outputs for security smells, and compared the results against 119 human-written Ansible roles from community repositories.
\subsubsection{Model Selection.}Table~\ref{ai_models_performance_comparison} presents the 16 AI models evaluated, spanning organisations including Anthropic, OpenAI, Google, Meta, and DeepSeek. The selection captures diversity across licensing models, context window sizes (128K to 10M tokens), and coding benchmark performance.\footnote{SWE-bench, \href{https://www.swebench.com/}{https://www.swebench.com/}.}\footnote{LiveCodeBench, \href{https://livecodebench.github.io/}{https://livecodebench.github.io/}.} All models were evaluated with standalone capabilities, disabling web search and retrieval augmented generation to ensure fair comparison.
\vspace{-5pt}
\begin{table}[!ht]
    \centering
    \caption{Features and Benchmark Performance Comparison of Large Language Models}
    \vspace{-8pt}
    \label{ai_models_performance_comparison}
    \resizebox{1\textwidth}{!}{%
    \normalsize
    \begin{tblr}{
      colspec = {|Q[3cm,c]|Q[3cm,c]|Q[2.5cm,c]|Q[3cm,c]|Q[1.8cm,c]|Q[1.75cm,c]|Q[3cm,c]|Q[5.25cm,l]|},
      row{1} = {font=\bfseries, bg=boxFrameColor, fg=white},
      cell{2-17}{1} = {bg=boxFrameColor!10, font=\bfseries},
      cell{2-17}{2-8} = {bg=rowColor2},
      vlines,
      hlines,
      vline{1} = {2.5pt, solid},
      vline{Z} = {2.5pt, solid},
      hline{1} = {2.5pt, solid},
      hline{Z} = {2.5pt, solid},
      hline{2} = {2.5pt, solid},
      hline{3-17} = {1.5pt, solid},
      vline{2-8} = {1.5pt, solid}
    }
    Organisation & Model Name & Parameters & Context Window & License & SWE Bench & LiveCodeBench & Benchmark Note \\
    Alibaba & Qwen3-235B-A22B & 235B & 128K & $\circ$ & \textcolor{red!60!black}{\xmark} & 70.70\% & LiveCodeBench v5 \\
    Amazon & Nova Pro & \textcolor{red!60!black}{\xmark} & 300K & $\bullet$ & 42.4\% & \textcolor{red!60!black}{\xmark} & SWE-bench Verified \\
    Anthropic & Claude Sonnet 4 & \textcolor{red!60!black}{\xmark} & 200K & $\bullet$ & 80.20\% & \textcolor{red!60!black}{\xmark} & SWE-bench high \\
    Anthropic & Claude Opus 4 & \textcolor{red!60!black}{\xmark} & 200K & $\bullet$ & 79.40\% & \textcolor{red!60!black}{\xmark} & SWE-bench high \\
    Anthropic & Claude Sonnet 3.7 & \textcolor{red!60!black}{\xmark} & 200K & $\bullet$ & 70.30\% & \textcolor{red!60!black}{\xmark} & SWE-bench high compute \\
    DeepSeek & DeepSeek V3~ & 671B & 128K & $\circ$ & 42.00\% & 37.60\% & SWE-bench Verified, LiveCodeBench chat model \\
    Google & Gemini 2.5 Pro & \textcolor{red!60!black}{\xmark} & 1M & $\bullet$ & 63.20\% & 75.60\% & SWE-bench Verified, LiveCodeBench v5 \\
    Meta & Llama 4 & 109B & 10M & $\circ$ & \textcolor{red!60!black}{\xmark} & 43.40\% & LiveCodeBench Maverick variant \\
    Mistral & Pixtral Large & 124B & 128K & $\bullet$ & \textcolor{red!60!black}{\xmark} & \textcolor{red!60!black}{\xmark} & \textcolor{red!60!black}{\xmark} \\
    Moonshot AI & Kimi K2 & 1T & 128K & $\circ$ & 65.80\% & 53.70\% & SWE-bench single-attempt, LiveCodeBench v6 \\
    OpenAI & GPT o3 & \textcolor{red!60!black}{\xmark} & 200K & $\bullet$ & 69.10\% & \textcolor{red!60!black}{\xmark} & SWE-bench Verified \\
    OpenAI & GPT o4-mini & \textcolor{red!60!black}{\xmark} & 200K & $\bullet$ & \textcolor{red!60!black}{\xmark} & \textcolor{red!60!black}{\xmark} & \textcolor{red!60!black}{\xmark} \\
    OpenAI & GPT 4o & \textcolor{red!60!black}{\xmark} & 128K & $\bullet$ & 33.20\% & \textcolor{red!60!black}{\xmark} & SWE-bench Verified \\
    OpenAI & GPT 4.1 & \textcolor{red!60!black}{\xmark} & 1M & $\bullet$ & 54.60\% & 44.70\% & SWE-bench Verified \\
    Perplexity & Sonar & \textcolor{red!60!black}{\xmark} & 127K & $\bullet$ & \textcolor{red!60!black}{\xmark} & \textcolor{red!60!black}{\xmark} & \textcolor{red!60!black}{\xmark} \\
    X (xAI) & Grok 4 & \textcolor{red!60!black}{\xmark} & 256K & $\bullet$ & $\sim$72\%--75\% & \textcolor{red!60!black}{\xmark} & Claimed by xAI \\
    \end{tblr}
    }
\small
\textit{Note:} \textcolor{red!60!black}{\xmark} indicates data not available or not disclosed; \textbf{License:} $\bullet$ = Closed Source,\\$\circ$ = Open Source; \textbf{Parameters:} B = Billion, T = Trillion; \textbf{Context Window:} K = Thousand tokens, M = Million tokens.
\vspace{-22pt}
\end{table}
\subsubsection{Tasks \& Evaluation Procedure:}Each model received an identical zero-shot prompt without examples or security guidance. This evaluates whether AI models produce secure code by default. The interaction comprised three phases: (Q1) a generation request: ``\textit{Generate an Ansible role for Tomcat 10 that installs and configures Host Manager for web applications}''; (Q2) a self-assessment query: ``\textit{Does the generated code follow any ISO code quality standards or best coding practices? (Yes/No)}''; and (Q3) a specification request if Q2 indicated adherence.

Generated code underwent quality analysis through the IaC quality framework~\cite{konala2025framework}, which assesses infrastructure code across nine dimensions including security and structure. The framework uses configurable policies to evaluate code against technology-specific standards. For this evaluation, we applied the Ansible best practices policy to measure AI-generated code against Ansible documentation guidelines. This framework uses quality scores to detect supply chain vulnerabilities~\cite{pandumeta} and hidden vulnerability propagation pathways~\cite{pandudiffu} that lie beyond the scope of IaC static analysis tools. Our evaluation extended beyond code analysis to examine AI models' awareness of quality standards through Q2 and Q3 responses. These self-reported responses may not reflect actual knowledge, as models can hallucinate standards adherence. However, comparing claimed awareness against actual code quality reveals gaps in model capabilities.
\subsubsection{Dataset Composition:}The evaluation dataset comprises 135 Ansible roles for Apache Tomcat 10: 16 AI-generated and 119 human-written roles from Ansible Galaxy~\footnote{Red Hat, Inc., ``Ansible Galaxy,'' 2016, \href{https://galaxy.ansible.com/}{https://galaxy.ansible.com/}.}. Tomcat was selected for two reasons: CIS Apache Tomcat 10 Benchmarks\footnote{Center for Internet Security, ``CIS Apache Tomcat v10 Benchmark,'' Version 1.1.0, February 2025, \href{https://www.cisecurity.org/benchmark/apache_tomcat}{https://www.cisecurity.org/benchmark/apache\_tomcat}.} containing 21 Level 1 security controls enable compliance assessment in subsequent sections, and Ansible Galaxy contains sufficient human-written implementations for comparison. The IaC quality framework produces scores on a 9-point scale based on Ansible documentation requirements.
\vspace{-0pt}
\subsection{Baseline Findings}
Evaluation produced dataset means of 4.56 for total quality (9-point scale, higher indicating better quality) and 0.634 for security (0--1 scale, higher indicating fewer security smells). Among 16 AI models, only Claude Opus 4 (5.07) exceeded the quality mean and only Perplexity Sonar (0.669) exceeded the security mean, with no model exceeding both. As Table~\ref{tab:security-quality-by-category} shows, AI-generated code consistently underperformed human-written implementations.
\begin{table}[!ht]
    \centering
    \caption{Security and Quality Performance Analysis by Standards Awareness Category}
    \vspace{-8pt}
    \label{tab:security-quality-by-category}
    \resizebox{1\textwidth}{!}{%
    \begin{tblr}{
      colspec = {|Q[2.25cm,c]|Q[3.25cm,l]|Q[7.0cm,l]|Q[2.0cm,c]|Q[2.25cm,c]|Q[2.25cm,c]|},
      row{1} = {font=\bfseries, bg=boxFrameColor, fg=white},
      cell{2-8}{1} = {bg=boxFrameColor!10, font=\bfseries},
      cell{2-8}{2} = {bg=green!20, font=\bfseries},
      cell{2-8}{3-6} = {bg=green!10},
      cell{9-14}{1} = {bg=boxFrameColor!10, font=\bfseries},
      cell{9-14}{2} = {bg=yellow!20, font=\bfseries},
      cell{9-14}{3-6} = {bg=yellow!10},
      cell{15-17}{1} = {bg=boxFrameColor!10, font=\bfseries},
      cell{15-17}{2} = {bg=red!20, font=\bfseries},
      cell{15-17}{3-6} = {bg=red!10},
      vlines,
      hlines,
      vline{1} = {2.5pt, solid},
      vline{Z} = {2.5pt, solid},
      hline{1} = {2.5pt, solid},
      hline{Z} = {2.5pt, solid},
      hline{2} = {2.5pt, solid},
      hline{3-17} = {1.5pt, solid},
      vline{2-6} = {1.5pt, solid},
      hline{9,15} = {2.0pt, solid},
    }
    Category & Model Name & Q3 Answer & Security Score (SS) & Total Quality Score (QS) & Category Avg (SS, QS) \\
    \SetCell[r=7]{c} Appropriate Standards & Claude Opus 4 & Ansible Best Practices & 85 (0.6136) \textcolor{declineRed}{$\downarrow$} & 21 (5.07) \textcolor{blue!100}{$\uparrow$} & \SetCell[r=7]{c} 0.599, 4.08 \\
     & GPT 4o & ISO/IEC 25010 \& 29110, OWASP IaC Sec Guide & 114 (0.6068) \textcolor{declineRed}{$\downarrow$} & 115 (4.06) \textcolor{declineRed}{$\downarrow$} & \\
     & DeepSeek V3 & ISO/IEC 25010, ISO/IEC 27001 (Indirect) & 122 (0.5944) \textcolor{declineRed}{$\downarrow$} & 117 (4.030) \textcolor{declineRed}{$\downarrow$} & \\
     & Grok 4 & General ISO/IEC 25010 Guidelines & 116 (0.6049) \textcolor{declineRed}{$\downarrow$} & 118 (4.01) \textcolor{declineRed}{$\downarrow$} & \\
     & Pixtral Large & Ansible Best Practices & 124 (0.5494) \textcolor{declineRed}{$\downarrow$} & 128 (3.85) \textcolor{declineRed}{$\downarrow$} & \\
     & Claude Sonnet 3.7 & General Software \& Ansible Best Practices & 84 (0.6142) \textcolor{declineRed}{$\downarrow$} & 129 (3.80) \textcolor{declineRed}{$\downarrow$} & \\
     & Qwen3-235B-A22B & Ansible Best Practices & 110 (0.6105) \textcolor{declineRed}{$\downarrow$} & 130 (3.73) \textcolor{declineRed}{$\downarrow$} & \\
    \SetCell[r=6]{c} No Standards & Claude Sonnet 4 & \textcolor{red!60!black}{\xmark} & 82 (0.6160) \textcolor{declineRed}{$\downarrow$} & 75 (4.45) \textcolor{declineRed}{$\downarrow$} & \SetCell[r=6]{c} 0.618, 3.95 \\
     & GPT o4-mini & \textcolor{red!60!black}{\xmark} & 114 (0.6068) \textcolor{declineRed}{$\downarrow$} & 116 (4.037) \textcolor{declineRed}{$\downarrow$} & \\
     & Kimi K2 & \textcolor{red!60!black}{\xmark} & 111 (0.6086) \textcolor{declineRed}{$\downarrow$} & 121 (3.99) \textcolor{declineRed}{$\downarrow$} & \\
     & Sonar & \textcolor{red!60!black}{\xmark} & 34 (0.6685) \textcolor{blue!100}{$\uparrow$} & 127 (3.88) \textcolor{declineRed}{$\downarrow$} & \\
     & GPT 4.1 & \textcolor{red!60!black}{\xmark} & 111 (0.6086) \textcolor{declineRed}{$\downarrow$} & 131 (3.72) \textcolor{declineRed}{$\downarrow$} & \\
     & Gemini 2.5 Pro & \textcolor{red!60!black}{\xmark} & 121 (0.5988) \textcolor{declineRed}{$\downarrow$} & 132 (3.65) \textcolor{declineRed}{$\downarrow$} & \\
    \SetCell[r=3]{c} Inappropriate Standards & Nova Pro & Confidential & 124 (0.5494) \textcolor{declineRed}{$\downarrow$} & 133 (3.61) \textcolor{declineRed}{$\downarrow$} & \SetCell[r=3]{c} 0.586, 3.57 \\
     & GPT o3 & PEP 8 for Ansible & 119 (0.6025) \textcolor{declineRed}{$\downarrow$} & 134 (3.58) \textcolor{declineRed}{$\downarrow$} & \\
     & Llama 4 & YAML standards for Ansible & 116 (0.6049) \textcolor{declineRed}{$\downarrow$} & 135 (3.51) \textcolor{declineRed}{$\downarrow$} & \\
    \end{tblr}
    }
\textbf{Legend:} Rank (Score) format among 135 Tomcat roles; Security (0--1) and Quality (9-point) scores, higher is better; \textcolor{declineRed}{$\downarrow$} below mean, \textcolor{blue!100}{$\uparrow$} above mean. \textbf{Colour:} \textcolor{green!90!black}{$\blacksquare$} Appropriate standards; \textcolor{yellow!90!black}{$\blacksquare$} No standards (\textcolor{red!60!black}{\xmark}); \textcolor{red!80!black}{$\blacksquare$} Inappropriate standards.
\end{table}

Categorisation based on Q2 and Q3 responses reveals three groups. These categories reflect claimed standards awareness rather than verified knowledge; models may cite standards without genuine understanding. Seven models citing domain-specific standards (Ansible Best Practices, ISO/IEC 25010) achieved quality scores averaging 4.08 but security scores of 0.599. Six models reporting no standards adherence achieved lower quality (3.95) but higher security (0.618). Three models citing inappropriate standards (PEP 8 for YAML, Ansible) produced both the lowest quality (3.57) and security (0.586) scores, indicating that misaligned guidance from non-applicable frameworks correlates with lower security and overall code quality.

Manual inspection revealed consistent security smells linked to Common Weakness Enumerations (CWEs) across AI-generated code. No model implemented Ansible Vault for credential management, exposing hardcoded sensitive data (CWE-798).\footnote{MITRE, ``CWE-798: Use of Hard-coded Credentials,'' \href{https://cwe.mitre.org/data/definitions/798.html}{https://cwe.mitre.org/data/definitions/798.html}.} Only Claude Opus 4 implemented error handling, with the remaining 15 models omitting this practice (CWE-755).\footnote{MITRE, ``CWE-755: Improper Handling of Exceptional Conditions,'' \href{https://cwe.mitre.org/data/definitions/755.html}{https://cwe.mitre.org/data/definitions/755.html}.} Six models failed to configure file permissions (CWE-276), namely Grok 4, Pixtral Large, GPT 4.1, Gemini 2.5 Pro, Nova Pro, and Llama 4.\footnote{MITRE, ``CWE-276: Incorrect Default Permissions,'' \href{https://cwe.mitre.org/data/definitions/276.html}{https://cwe.mitre.org/data/definitions/276.html}.} No model included inline comments, and only 3 models (Claude Opus 4, Grok 4, and Sonar) included README files. Structural compliance was higher, with all models following directory conventions and 10 out of 16 producing syntax-error-free code. Grok 4, Qwen3-235B-A22B, GPT 4.1, Nova Pro, GPT o3, and Llama 4 contained syntax errors.

These findings address \textbf{RQ1}. The default security posture of AI-generated IaC is inadequate, with all 16 models producing code containing security smells and consistently underperforming human-written implementations. Structural correctness masks security violations including hardcoded credentials, missing error handling, and absent access controls. This baseline does not provide a foundation for compliance hardening, as regulatory frameworks presuppose security foundations that AI-generated code lacks by default. This inadequacy motivates \textbf{RQ2}, whether security-compliance requirements can be embedded into the generation process rather than remediated post-synthesis.
\section{Methodology}
To address RQ2, we present compliance-guided generation, an approach that embeds Ansible best practices and regulatory requirements as constraints during synthesis rather than detecting violations post-generation. Because baseline evaluation showed that models claim standards awareness without acting on it, our methodology makes the meaning of quality and compliance explicit rather than assuming the model already applies it. This requires mapping regulatory requirements to measurable code security attributes. The IaC code quality framework~\cite{konala2025framework}, applied earlier in that evaluation, provides these measurements through repository-wide analysis. This investigation employs CIS Benchmarks rather than DISA STIGs, as CIS targets commercial systems across various industries, as established in Section~\ref{cis_discussion}, whereas STIGs focus on federal government deployments.
\begin{figure}[!ht]
\centering
\includegraphics[width=1\textwidth]{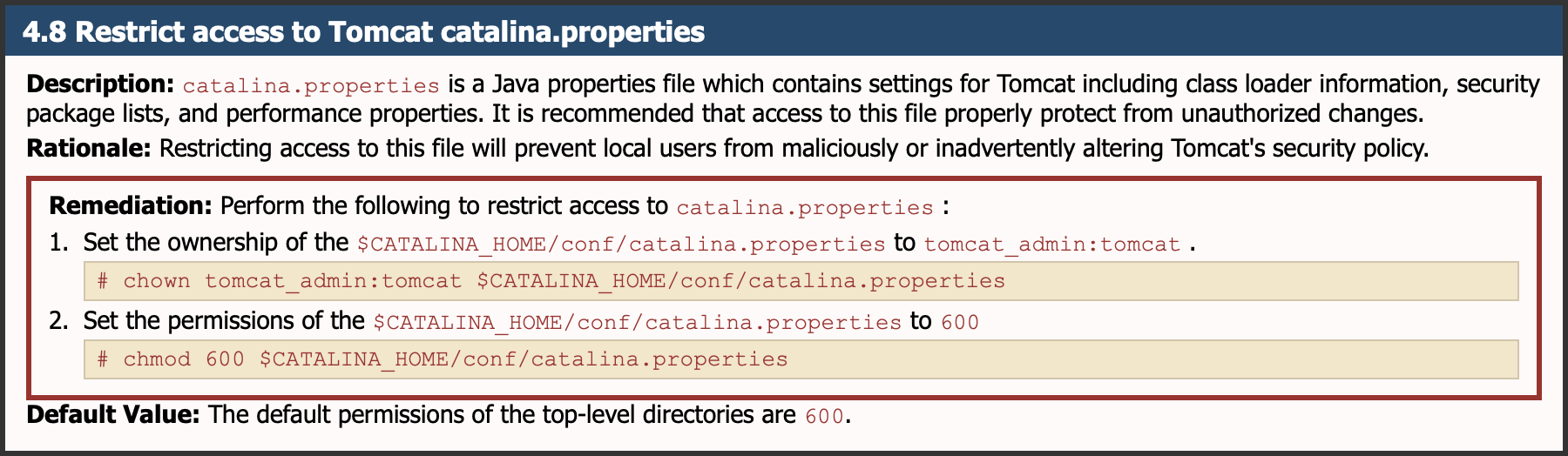}
\vspace{-20pt}
\caption{Sample CIS Level 1 (IG1) Control from Apache Tomcat 10 Benchmark}
\label{Figure:sample_cis}
\vspace{-15pt}
\end{figure}

Figure~\ref{Figure:sample_cis} presents a CIS Level 1 security control from the Apache Tomcat 10 Benchmark, illustrating how compliance requirements specify concrete technical security configurations. The remediation specifications show how requirements translate to code security attributes: \texttt{chown} and \texttt{chmod} commands specify ownership and permissions mapping to security attributes, while directory path specifications map to code structure attributes such as configuration templates.

We map CIS attributes to two dimensions of the IaC quality framework, which aligns with ISO/IEC 25010 (Software Quality Model). Structure scores serve as a primary filter: technology-specific compliance requirements demand configuration template files (e.g., \texttt{catalina.properties} for Tomcat, \texttt{mongod.conf} for MongoDB) where security controls are implemented. Without these templates, security controls cannot be applied, regardless of other code properties. Code security scores validate content within these templates, as compliance requirements specify permission values and configuration strings mapping to security attributes.
\begin{figure}[!ht]
\centering
\includegraphics[width=1\textwidth]{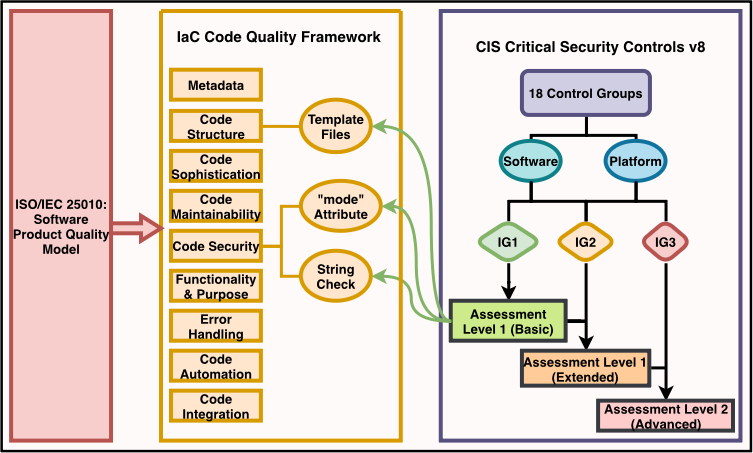}
\vspace{-20pt}
\caption{Alignment of CIS Critical Security Controls (Level 1) with IaC Code Quality Framework Components}
\label{Figure:integration_comp_quality}
\vspace{-15pt}
\end{figure}
Figure~\ref{Figure:integration_comp_quality} illustrates this mapping from CIS security controls to framework categories. Roles achieving high scores in both dimensions indicate compliance readiness. Post-deployment validation through CIS-CAT Pro supports correlation between framework-based assessment and regulatory benchmark outcomes.
\vspace{-15pt}
\subsection{The Extended CO-STAR Prompting Framework}
With assessment mechanisms established, we require a method to guide AI models toward generating quality-compliant code. The CO-STAR framework~\cite{costar-framework} structures generation instructions through six components: \textbf{C}ontext defines model role and expertise domain, \textbf{O}bjective specifies task requirements, \textbf{S}tyle establishes code formatting conventions, \textbf{T}one determines documentation characteristics, \textbf{A}udience identifies target users, and \textbf{R}esponse defines expected output format. We adopt this framework for a zero-shot approach to generating compliant code. Unlike Chain-of-Thought~\cite{wei2023chainofthoughtpromptingelicitsreasoning} or Tree-of-Thoughts~\cite{yao2023treethoughtsdeliberateproblem} requiring multiple interactions and API calls, or few-shot learning demanding curated example sets, zero-shot generation requires only a single interaction. This approach reduces experimental variables from example selection or iterative refinement, enabling direct comparison across AI models for generalisable results.

We extended CO-STAR with two categories for adding security requirements. The QUALITY category embeds Ansible best practices such as error handling and credential management through Ansible Vault. The COMPLIANCE category embeds CIS security controls specifying service configurations and access restrictions. Listing~\ref{costar_quality_compliance} demonstrates this structure. Each requirement specifies an attribute type (\texttt{mode} for permissions, \texttt{string} for configuration values, \texttt{readme\_file} for documentation), an objective from Ansible best practices or CIS documentation, and an implementation directive. 
\begin{center}    
\begin{promptbox}[label={costar_quality_compliance}]{Sample of Extended CO-STAR Prompt}
\begin{lstlisting}

CONTEXT: You are a senior DevOps engineer and Ansible expert specializing in infrastructure as code. Read and understand all the Steps below carefully before proceeding.
...
STEP 1: [INSTRUCTIONS]
Read and understand all instructions below carefully before proceeding.
OBJECTIVE: Give me an ansible role for tomcat which installs and configures Host Manager for webapps.
STYLE: Professional, avoid feature bloat, use bare minimum code, and stay LEAN - prioritise clarity over comprehensiveness.
TONE: Clear and technical with comprehensive examples
AUDIENCE: DevOps engineers, system administrators and interns
RESPONSE: Complete Ansible playbook structure with validation and testing
CONDITIONS: 'A' represents that it is a mandatory requirement, 'B' represents it is a recommended but context dependent. 'C' represents anti-patterns which should not be used.
STEP 2: [TECHNICAL REQUIREMENTS]
Generate the Ansible code strictly following these requirements based on their CONDITIONS values and avoid feature bloat. FAILURE WILL REQUIRE REGENERATION:
\end{lstlisting}
\begin{lstlisting}[firstnumber=13,backgroundcolor=\color{yellow!40}]

** COMPLIANCE:
-mode=A:Restrict access to Tomcat logs directory->Set the ownership of the $CATALINA_HOME/logs to tomcat_admin:tomcat and permissions to 0750.
-string=A:Disable the Shutdown port->Set the port to -1 in the $CATALINA_HOME/conf/server.xml to disable the shutdown port
** QUALITY:
-readme_file=A: Generate comprehensive README.md
-debug=B: Include debug tasks for troubleshooting
-password=C: Avoid usage of password
\end{lstlisting}
\begin{lstlisting}[firstnumber=20]
STEP 3: [VERIFICATION]
Recheck and confirm all TECHNICAL REQUIREMENTS are met. FAILURE WILL REQUIRE REGENERATION and return to STEP 1 and REDO.
\end{lstlisting}
\end{promptbox}
\end{center}

Requirements use priority classification. Category `A' denotes mandatory requirements from both Ansible best practices and CIS controls. Category `B' denotes context-dependent practices. Category `C' denotes anti-patterns to avoid, including hardcoded passwords and insecure defaults. The verification step instructs models to validate generated code against all requirements before completion.

\subsection{Dataset Overview}
To evaluate the extended CO-STAR framework, we expanded the dataset beyond Tomcat to include MongoDB, enabling investigation across technologies with varying security control complexity. Human-written roles were sourced from Ansible Galaxy, the official community repository providing community-validated implementations representing real-world deployment practices. The complete dataset comprises 278 Ansible roles:
\begin{itemize}
\item 135 Tomcat roles (16 AI-generated, 119 human-written) validated against 21 CIS Level 1 controls; input prompt comprises 3,403 tokens (1,696 words)
\item 143 MongoDB roles (16 AI-generated, 127 human-written) assessed against 7 CIS Level 1 controls;\footnote{Center for Internet Security, ``CIS MongoDB v7 Benchmark,'' Version 1.2.0, August 2025, \href{https://www.cisecurity.org/benchmark/mongodb}{https://www.cisecurity.org/benchmark/mongodb}.} input prompt comprises 1,572 tokens (873 words)
\end{itemize}
MongoDB was selected for two reasons. Human-written ansible role counts (127) align with Tomcat (119), facilitating comparative analysis. Its 7 security controls contrast with Tomcat's 21, testing whether models maintain security accuracy as requirement count increases. This investigation focuses on CIS Level 1 controls establishing baseline security. The same 16 models from baseline evaluation were used. All roles were tested by deploying them in isolated test environments, with post-deployment validation performed using CIS-CAT Pro for assessing security compliance levels achieved.
\vspace{-5pt}
\section{Results and Discussion}
Of the 16 evaluated models, only four generated syntactically correct code satisfying quality and compliance constraints: Claude Opus 4, Claude Sonnet 3.7, Gemini 2.5 Pro, and Pixtral Large. Compliance success correlated with standards awareness from baseline evaluation: three successful models came from the appropriate standards category, one from the no standards category, and none from the inappropriate standards category.

This 75\% failure rate occurred despite input prompt sizes (1,572--3,403 tokens) representing less than 3\% of the smallest context window (128K tokens) among evaluated models.\footnote{IBM Research, ``Larger Context Window,'' \href{https://research.ibm.com/blog/larger-context-window}{https://research.ibm.com/blog/larger-context-window}.} Claude Sonnet 4 achieved the highest SWE-bench score (80.2\%) yet failed due to incomplete code and syntax errors, while Pixtral Large succeeded despite lacking benchmark data in popular assessment platforms shown in Table~\ref{ai_models_performance_comparison}. Similarly, Llama 4 with 10M tokens failed while Claude Opus 4 with 200K tokens succeeded. These results suggest that instruction-following capability for multi-constraint security tasks likely emerges from training data rather than architectural features alone.

The extended CO-STAR framework functions as an instruction-following test requiring models to parse multi-layered prompt structure, understand categorical requirement systems (A/B/C priority classification), interpret attribute-specific syntax (\texttt{mode=A:}, \texttt{password=C:}), and apply constraints across quality and compliance domains. The four successful models are closed-source systems, preventing further investigation into underlying success factors without collaboration with model providers.
\begin{table}[!ht]
    \centering
    \caption{AI Models Pre-deployment Quality Scores and Post-deployment Compliance}
    \vspace{-5pt}
    \label{ai_models_compliance_results}
    \resizebox{1.0\textwidth}{!}{%
    \begin{tblr}{
      colspec = {|Q[3.0cm,l]|Q[1.5cm,c]|Q[1.8cm,c]|Q[1.8cm,c]|Q[2cm,c]|Q[2cm,c]|Q[1.8cm,c]|Q[1.8cm,c]|},
      row{1-2} = {font=\bfseries, bg=boxFrameColor, fg=white},
      cell{3-5}{1} = {bg=green!20, font=\bfseries},
      cell{3-5}{2} = {bg=gray!10},
      cell{3-5}{3,5,7} = {bg=tomcatColor},
      cell{3-5}{4,6,8} = {bg=mongoBlue},
      cell{6}{1} = {bg=yellow!20, font=\bfseries},
      cell{6}{2} = {bg=gray!10},
      cell{6}{3,5,7} = {bg=tomcatColor},
      cell{6}{4,6,8} = {bg=mongoBlue},
      vlines,
      hlines,
      vline{1} = {2.5pt, solid},
      vline{Z} = {2.5pt, solid},
      hline{1} = {2.5pt, solid},
      hline{Z} = {2.5pt, solid},
      hline{2} = {1.5pt, solid},
      hline{3} = {2.5pt, solid},
      hline{4-5} = {1.5pt, solid},
      hline{6} = {2.0pt, solid},
      vline{2-8} = {1.5pt, solid}
    }
    \SetCell[r=2]{c} Model & \SetCell[r=2]{c} Context Window & \SetCell[c=2]{c} Structure/Security Score & & \SetCell[c=2]{c} CIS-CAT Compliance & & \SetCell[c=2]{c} Rank (Total Quality Score) & \\
    & & Tomcat & MongoDB & Tomcat & MongoDB & Tomcat & MongoDB \\
    Claude Opus 4 & 200K & 1.0/0.94 & 1.0/1.0 & 95.2\% & 100\% & 1 (6.03) & 3 (5.61) \\
    Claude Sonnet 3.7 & 200K & 1.0/0.85 & 1.0/0.85 & 85.7\% & 85.7\% & 8 (5.38) & 7 (5.37) \\
    Pixtral Large & 128K & 0.9/0.62 & 1.0/0.85 & 47.1\% & 85.7\% & 37 (4.91) & 50 (4.67) \\
    Gemini 2.5 Pro & 1M & 0.9/0.57 & 1.0/0.85 & 38.1\% & 85.7\% & 5 (5.45) & 27 (4.88) \\
    \end{tblr}
    }
\textbf{Colour Coding:} \textcolor{green!90!black}{$\blacksquare$} = Appropriate standards; \textcolor{yellow!90!black}{$\blacksquare$} = No standards
\end{table}

Table~\ref{ai_models_compliance_results} presents context window sizes, pre-deployment structure/security scores, post-deployment CIS-CAT compliance outcomes, and total quality scores for the four successful models, demonstrating that pre-deployment quality metrics indicate regulatory conformance. Analysis reveals correlation patterns: higher structure/security scores correspond to increased compliance rates, while MongoDB's reduced control count (7 vs. Tomcat's 21) enabled higher compliance achievement across models. These results establish that structure and security metrics can enable pre-deployment verification of compliance readiness.

Human-written Ansible Galaxy roles also demonstrated compliance failures. Among 119 Tomcat roles, 113 (95.0\%) lacked configuration templates for CIS security controls (structure score $<$ 1), while the six roles with templates achieved maximum 23.8\% compliance. For MongoDB, 118 of 127 roles (92.9\%) similarly lacked templates (structure score $<$ 1), with nine implementing roles achieving maximum 42.85\% compliance. Failures across both technologies resulted from incorrect permission syntax, malformed configuration strings, and missing security-compliance implementations. Users adopting these human-written roles for regulated deployments face substantial remediation effort, as even the highest-performing implementations require correction of 57.1\%--76.2\% of controls before achieving compliance authorisation and quality improvement. 

Beyond compliance, successful AI-generated roles achieved competitive quality rankings among all 278 implementations. Claude Opus 4 ranked 1st and 3rd (6.03, 5.61), Claude Sonnet 3.7 8th and 7th (5.38, 5.37), Gemini 2.5 Pro 5th and 27th (5.45, 4.88), and Pixtral Large 37th and 50th (4.91, 4.67) for Tomcat and MongoDB respectively. Manual inspection of these roles found none of the security smells identified during baseline analysis. All four models implemented Ansible Vault for credential management and included proper file permissions. Compared to baseline evaluation where models received no security guidance (Table~\ref{tab:security-quality-by-category}), quality scores improved by 19\%--49\%. Claude Opus 4 improved from 5.07 to 6.03 (19\%), Pixtral Large from 3.85 to 4.91(28\%), Claude Sonnet 3.7 from 3.80 to 5.38(42\%), and Gemini 2.5 Pro from 3.65 to 5.45(49\%). For the four successful models, structured prompting with quality and compliance constraints produced code free of security smells, providing a strong foundation for compliance whilst achieving higher code quality than human-written implementations.

For capable models, quality-compliance requirements can be embedded into code generation, eliminating iterative post-generation remediation. The leading model achieved 95.2\%--100\% CIS compliance through single-generation structured prompting compared to 23\%--43\% for human-written code. The extended CO-STAR framework demonstrates that quality-compliance guided generation produces results that detection-based approaches cannot attain, addressing \textbf{RQ2}.
\subsection{Prevention Versus Detection}
Current approaches to IaC security rely on detection, with static analysis tools identifying security smells after code enters repositories. This paradigm is inadequate for AI-generated code because it cannot prevent security smells from being introduced during synthesis, instead requiring iterative remediation cycles that incur time and computational costs. Our findings support a prevention-based alternative grounded in the observation that code quality serves as a prerequisite for compliance. CIS controls assume that foundational code is already of high quality with proper security measures such as permissions and credential management in place, providing the base upon which compliance controls are applied. For capable models, embedding quality-compliance controls during synthesis produces compliant code rather than requiring post-hoc correction. The correlation between pre-deployment quality metrics and post-deployment CIS-CAT outcomes further validates this approach, enabling compliance verification before code reaches production environments. Prevention-based generation complements rather than replaces existing static analysis tools, as detection remains valuable for auditing committed code while quality-compliance guided code generation prevents violations from entering codebases in the first instance.
\subsection{Implications for Practitioners}
Compliance-aware generation improves program comprehension for developers and the IaC community. Our methodology serves as executable documentation, with security requirements explicitly stating objectives alongside implementation directives, making security intent transparent in the codebase. Code generated by successful models exhibits consistent structure with configuration templates, proper directory organisation, and README files, reducing cognitive load when onboarding team members or auditing deployments. The mapping between CIS controls and code attributes provides traceability from security requirements to implementation, enabling developers to understand why specific configurations exist. This contrasts with human-written implementations where security rationale is rarely documented, forcing maintainers to reverse-engineer intent from configuration values. For the IaC community, our extended CO-STAR prompts establish reusable patterns encoding security expertise, lowering the barrier for practitioners unfamiliar with best practices and regulatory frameworks. These prompts integrate with AI models and coding assistants as system prompts or can function independently as user prompts.
\subsection{Limitations and Future Work}
\textbf{Internal validity:} Our evaluation captures AI models capabilities at a specific point in time; AI models evolve rapidly, and results may differ with newer versions which is a limitation common to empirical studies in this rapidly evolving domain. \textbf{External validity:} The evaluation focuses on two technologies (Apache Tomcat and MongoDB) with CIS Level 1 controls using Ansible and zero-shot interactions as a foundational study. Dataset composition leverages publicly available Ansible Galaxy roles representing community contributions that may differ from proprietary organisational deployments. \textbf{Construct validity:} The mapping from IaC quality framework dimensions to CIS compliance requirements assumes structure and security scores indicate compliance readiness; while post-deployment CIS-CAT validation supports this correlation, independent validation across diverse regulatory frameworks remains necessary.

In future work, we will expand this investigation to higher implementation groups (IG2, IG3), alternative compliance frameworks (STIG, PCI-DSS), diverse IaC technologies (Terraform, Puppet), and multi-turn refinement approaches. Comparative analysis of fine-tuning versus prompting would quantify whether embedding compliance knowledge through model training yields superior outcomes. Fine-tuning open-source models using Low-Rank Adaptation (LoRA)~\cite{hu2021lora} on compliant IaC code could internalise security practices and compliance requirements, potentially addressing instruction-following limitations observed in the 75\% of models that failed complex multi-constraint tasks.
\vspace{-15pt}
\section{Conclusion}
\vspace{-10pt}
This paper investigated whether AI models can generate quality-compliant IaC. Evaluation of 16 AI models and 278 Ansible roles demonstrated three findings. First, all 16 AI models produced code containing security smells by default, underperforming human-written implementations. Second, 12 out of 16 models failed not from coding inability but from instruction-following limitations with multiple constraints. Models achieving 80\% on SWE-bench failed while others lacking benchmark data succeeded. Third, structured prompting enables four models to generate compliant code, with the leading model achieving 95\%--100\% CIS compliance compared to 23\%--43\% for human implementations, with overall code quality improving by 19\%--49\%. Manual inspection confirmed the four successful roles contained no security smells previously detected, providing a strong foundation for compliance. All successful models are closed-source, indicating secure-compliant generation depends on capabilities emerging from training methodologies rather than architectural features alone. For capable models, our approach deploys immediately through system prompts without model retraining. Validation across higher implementation groups, alternative frameworks, and diverse IaC technologies remains necessary to establish generalisability.
\vspace{-15pt}
\section{Data Availability}
\vspace{-10pt}
The dataset and prompts of this study are available at: \url{https://figshare.com/s/67c36294c704c7ab5ae5}.
\vspace{-10pt}
\bibliographystyle{splncs04}
\bibliography{ref}
\end{document}